# Giant Hyperfine Interaction between a Dark Exciton Condensate and Nuclei

A. Jash[1,*], M. Stern[2], S. Misra[1], V. Umansky[1], and I. Bar Joseph[1]

[1]*Department of Condensed Matter physics, Weizmann Institute of Science, Rehovot 7610001, Israel*
[2]*Department of Physics, Bar-Ilan University, Ramat-Gan 5290002, Israel*

Abstract

We study the interaction of a dark exciton Bose-Einstein condensate with the nuclei in GaAs/AlGaAs coupled quantum wells and find clear evidence for nuclear polarization buildup that accompanies the appearance of the condensate. We show that the nuclei are polarized throughout the mesa area, extending to regions which are far away from the photoexcitation area, and persisting for seconds after the excitation is switched off. Photoluminescence measurements in the presence of RF radiation reveal that the hyperfine interaction between the nuclear and electron spins is enhanced by two orders of magnitude. We show that this large enhancement manifests the collective nature of the $N$-excitons condensate, which amplifies the interaction by a factor of $\sqrt{N}$.

---

[*] amit-kumar.jash@weizmann.ac.il.

Excitons in semiconductors are expected to undergo a Bose–Einstein condensation (BEC) transition at low temperatures [1-3]. Coupled quantum wells (CQW) or bilayer structures, where electrons and holes are confined to distinct potential wells, have emerged as a leading platform for realizing this transition [4-24]. The small overlap between the electron and hole wavefunctions, which bind to form spatially indirect exciton, IX, results in an extended exciton lifetime and a diminished spin relaxation rate [25], and allow reaching the critical density for condensation at relatively low power.

A significant contribution to this endeavor, shedding light on the impact of the electron-hole exchange interaction, stems from the work of Combescot et al. [9, 12]. Their findings suggest that, due to this interaction, the optically inactive (spin-forbidden) dark exciton state is lower in energy relative to the optically active bright exciton, and is therefore the ground state of the condensate. Consequently, above the condensation threshold, excitons are anticipated to predominantly occupy this long-lived ground state. Unfortunately, however, the lack of light emission from the condensate makes it difficult to study the properties of this dark condensate and confirm its collective nature. Experimental efforts have therefore focused on studying the behavior of the bright IXs in the presence of the condensate, in particular, its blueshift above a certain threshold power and below a critical temperature [10]. This blueshift, $\Delta E$, which is due to the repulsive interaction between the IXs, was shown to manifest the buildup of a high exciton density in the dark ground state [17-20, 22, 24]. Other findings substantiating the occurrence of condensation take advantage of dark-bright exciton mixing, which turns the condensate to be partially radiative [12]. These include observation of spatial coherence [6, 7, 11, 17], spin polarization [18] and spin textures [14], defects proliferation representing quantized vortices [21], and enhanced luminescence [23].

In this work we study the condensate hyperfine interaction with the nuclei and show that it can be an effective probe that reveals its collective properties. It is well known that the hyperfine interaction of optically excited electrons with nuclei in semiconductors may transfer the electrons spin to the nuclei, in a process known as dynamical nuclear polarization (DNP) [26-28]. Numerous studies in the past two decades have demonstrated that the nuclei in a quantum dot can be effectively polarized by continuous resonant excitation with circularly polarized light [29-35]. Here we show that the continuous introduction of dark excitons, having a well-defined spin polarization, to the condensate ground state creates favorable conditions for DNP. The spin exchange (flip-flop) between an electron in the condensate ground state, $|G_N\rangle$, and a nucleus, turns a dark exciton into bright, while transferring the electron spin to the nucleus. The bright exciton in the excited state, $|E_N\rangle$, is quickly removed from the system by radiative recombination, reducing the total number of excitons from $N$ to $N-1$ (Fig. 1a). The net effect of this process is flipping the nuclear spin from ↑ to ↓, and repeating it by pumping more and more dark excitons we may create macroscopic nuclear polarization.

We investigate the PL and radio-frequency (RF) spectrum of optically excited GaAs/AlGaAs CQW at low temperatures and find clear evidence for DNP that accompanies the appearance of the dark condensate. We show that the nuclei are polarized throughout the area of the mesa, extending to regions which are a few hundred μm from the excitation beam, and persisting for a few seconds after the optical excitation is switched off. Remarkably, we find that the RF excitation of the condensate is in the tens MHz range, $\sim 10^2$ larger than the single exciton hyperfine splitting. We show that this large enhancement is proportional to $\sqrt{N}$, where $N$ is the number of excitons in the condensate, allowing us to determine the condensate coherence volume.

The CQW structure consists of an 18 nm wide well (WW) and a 12 nm narrow well (NW), separated by a 3 nm $Al_{0.3}$GaAs barrier. Top and bottom doped layers, which are 1 $\mu m$ above and below the CQW, allow the application of a gate voltage, $V_g$. Unlike many experiments on exciton BEC, which use an electrostatic trap or local disorder to confine the excitons in a small area, the samples we study in this work are large area mesa, $\sim 10^5$ $\mu m^2$ (Fig. 1b). We show below that this structure allows dark excitons to diffuse away from the excitation region to the entire area of the mesa and polarize the nuclei throughout.

By tuning the excitation laser energy below the NW energy gap, we generate electrons and holes solely in the WW. Some of the photo-excited electrons may tunnel to the NW due to the applied electric field and bind to the holes in the WW to form IX, while the remaining electrons in the WW bind to these holes and form direct excitons or positively charged trions. Indeed, the low power photoluminescence (PL) spectrum displays these three lines:

a direct exciton line at 1.524 eV (DX), a trion line 1meV below it (T), and a broad IX line further below, at an energy that depends on $V_g$ (Fig. 1c) [36].

**Dark Exciton Condensation and Nuclear Polarization**

We begin by presenting the evolution of the spectrum with power, showing that it exhibits a clear threshold behavior. Figure 2a shows a typical power dependence of the trion line intensity, $I_T$, and IX energy blueshift, $\Delta E$. It is seen that $I_T$ first grows linearly with power, and then sharply falls to zero at $P = 5\ \mu W$. The IX exhibits corresponding changes, steeply blueshifting when $I_T$ falls to zero (red curve in Fig. 2a). Since $\Delta E$ is proportional to the IX density, $n_{IX}$, the threshold power, $P_{th}$, at which it steeply rises signifies the point where $n_{IX}$ sharply increases. Clearly, the trion and IX compete on the same holes in the WW, and therefore the sharp increase of $n_{IX}$ necessarily implies a drop of $I_T$. Hence, we can use both the fall of $I_T$ and the increase of $\Delta E$ to determine $P_{th}$ (See SI section 4). We observe this behavior, namely, an abrupt fall of $I_T$ and steep rise of $\Delta E$ above a threshold power, over a broad range of gate voltages and temperatures.

We can unambiguously show that the steep rise of $\Delta E$ above $P_{th}$ is due to dark excitons density buildup. Figure 2b depicts the results of pump-probe measurement, where we excite the mesa at one end and measure the PL spectrum by a weak probe a few hundreds of µm away from the excitation spot. The spectrum at the probe location exhibits a strong dependence on pump power: $\Delta E$ sharply increases and $I_T$ falls when the pump power exceeds $P_{th}$, implying that the IXs created by the pump have a very long lifetime ($\sim 100's\ \mu s$) and can therefore diffuse over long distances [24].

Figure 2c shows the time evolution of $\Delta E$ and $I_T$ at the pump location, following an abrupt turn on of the excitation beam at a power level exceeding $P_{th}$. It is seen that the increase of $\Delta E$ and corresponding decrease of $I_T$ start after a long delay time, $\tau_d$, of ~1 second and reach their steady-state value after a few more seconds. We find that $\tau_d$ depends on the excess pump power above threshold, $\tau_d \propto 1/(P - P_{th})$, such that at $P \approx P_{th}$ it may reach tens of seconds (See SI section 4). This slow dynamic is found over a broad parameter range of gate voltages and in various mesas of different shapes and dimensions from the same wafer, indicating that it manifests an intrinsic process [16].

Such long response times in semiconductors are indicative of interaction with nuclear spins, which are known to relax at a very low rate through dipolar interaction with neighboring nuclei. Indeed, spin exchange that turns a dark exciton into bright is a *loss mechanism* for the condensate, which limits the occupation of the dark exciton ground state above threshold. As more and more nuclei become polarized with increasing time or pump power above threshold, this process is gradually quenched, and the dark exciton density may grow. Consequently, the rise time of the dark exciton density reflects the buildup time for the nuclear polarization to reach its steady state.

The open geometry, which allows dark excitons to diffuse away from the excitation region, implies that this nuclear polarization process occurs throughout the mesa. This is evident in the slow evolution of the probe spectrum in a pump-probe measurement. We find that the changes of $I_T$ and $\Delta E$ at the probe location appear only after a long delay time, which increases as the distance between the pump and probe beams grows (Fig. 2d). In section 2 of the SI, we present a simple one-dimensional diffusion model, which accounts for the DNP generated by the dark excitons and reproduces the observed slow dynamics.

**The Giant Hyperfine Interaction between the Condensate and Nuclei**

To obtain further insight into the role of nuclear polarization in determining the optical spectrum, we performed power-dependent PL measurements while exposing the sample to RF radiation. The underlying idea is that at certain frequencies in the RF range, this radiation can be resonantly absorbed by the nuclei, causing them to flip their spin and thereby affect $P_{th}$ (See SI section 1 for experimental details). In Fig. 3a we show $P_{th}$ as a function of RF, where each point is obtained by fixing the RF at a specific value and measuring the power dependence of the PL spectrum. Remarkably, we find pronounced dips in $P_{th}$ at three resonance frequencies, $f_i = 24, 39,$ and 52 MHz.

The impact of the RF radiation on the optical spectrum is explicitly demonstrated in Fig. 3b and 3c, which show $\Delta E$ and $I_T$ as a function of pump power with (black) and without (red) RF radiation at $f_1 = 24$ MHz. We find that the resonance relative depth, defined as $P_{th}(f_1)/P_{th}(0)$, is proportional to the RF power, and may reach two orders of magnitude reduction of $P_{th}$ at the high RF power limit (see SI section 4).

The existence of three RF resonances at zero magnetic field suggests that they are associated with transitions between hyperfine levels of the three isotopes in the GaAs system: $^{75}$As, $^{69}$Ga, and $^{71}$Ga. Indeed, the natural abundance of these isotopes, 100%, 60%, 40%, respectively [27, 37], matches the relative strength of the resonances. In the Fig. 3d we present the measured frequencies, $f_i$, as a function of the gyromagnetic ratio of the three isotopes ($\gamma_{As}^{75} = 7.29$, $\gamma_{Ga}^{69} = 10.22$, and $\gamma_{Ga}^{71} = 12.98 \times 10^6$ s$^{-1}$ T$^{-1}$ [37]). It is seen that there is a perfect linear relationship between the two, indicating that $f_i$ are proportional to $\gamma_i$. Surprisingly, however, the values of $f_i$ are very high, $\sim 10^7$ Hz, about two orders of magnitude larger than the $\sim 10^5$ Hz hyperfine splitting expected for a GaAs exciton [27]. We show below that this large enhancement of the electron-nuclei coupling is a manifestation of the collective interaction of the condensate and the nuclei.

Let us consider the collective hyperfine Hamiltonian, $H_{hf}$, of $N$ electrons in the exciton BEC ground state, each having $+s_z$ spin and coupled to a nucleus with a hyperfine coupling constant $g$. We can view the electrons as a collection of $N$ two-level systems, each having an energy gap of $g s_z I_z$ that depends on the nuclear spin, $I_z$. $H_{hf}$ can be expressed in terms of the total electron spin operator $\mathbf{S} = \sum_{j=1}^{N} \mathbf{s}^j$, as $H_{hf} = g[S_z I_z + S_+ I_- + S_- I_+]$. Since the system is in its BEC ground state, its collective wavefunction can be written as $|G_N, \uparrow\rangle = |0,0,0,0,\ldots 0\rangle$, and its collective excited state as a superposition of single spin excitations, $|E_N, \downarrow\rangle = 1/\sqrt{N}(|1,0,0,0\ldots,0\rangle + |0,1,0,0\ldots,0\rangle + |0,0,1,0\ldots,0\rangle + \ldots)$. Here the arrows ↑,↓ mark the nuclear spin orientation at the initial and final states, respectively. It is easy to see that the matrix element between the two states is $\langle E_N, \downarrow | H_{hf} | G_N, \uparrow \rangle = \frac{1}{\sqrt{N}} \sum_{j=1}^{N} g/2 = \sqrt{N} g/2$, and the energy difference between the collective ground and excited states is $\sqrt{N} g$, enhanced by a large factor of $\sqrt{N}$ relative to the single particle case (See SI section 3 for a detailed derivation). Accordingly, the $\sim 10^2$ enhancement factor implies that the number of excitons in the condensate is $N \approx 10^4 - 10^5$. Taking the density to be $\sim 10^{10}$ cm$^{-2}$ the coherence length can be estimated to be $\sim 10$ μm.

**The Hyperfine Interaction in a Magnetic field**

Support for this interpretation is provided by measurements in a magnetic field, $B_{ext}$, oriented along the $+z$ direction. The field breaks the degeneracy between the $|\pm 2\rangle$ dark exciton states, such that the electron polarization in the condensate ground state is $|+\frac{1}{2}\rangle$, and the energy gap to the excited state at threshold is the Zeeman energy, $\hbar\omega_e$. However, spin flip transitions from dark to bright excitons polarize the nuclei along the $+z$ direction, yielding an Overhauser field, $B_O$, on the electrons that is oriented opposite to $B_{ext}$ and closing the Zeeman gap [30]. As the pump power is increased, more and more nuclei are polarized, and $B_O$ grows until it compensates the external field, $B_O = -B_{ext}$ [38, 39]. Increasing the pump power beyond this value will not result in further polarization of the nuclei, and the dark exciton density may start growing.

The evolution of the spectrum in a magnetic field nicely shows this behavior. Figure 4a depicts $I_T$ and $\Delta E$ as a function of power for several magnetic fields. It is seen that the saturation power of $I_T$, which marks the condensation onset at $P \approx 2~\mu W$, is almost independent of magnetic field. However, it is evident that there is an intermediate broad power range following this onset, in which $I_T$ remains approximately constant. Only beyond this power range - $I_T$ falls to zero and $\Delta E$ exhibits a steep rise. This intermediate power range is the range at which the excess power goes to polarizing the nuclei. Accordingly, the power at which $I_T = 0$ and $\Delta E$ exhibits a steep rise is the point at which $B_O = -B_{ext}$.

The increase of $P_{th}$ in the range $0 < B_{ext} < 1.75$ T seen in Fig. 4a and 4b agrees with this predicted behavior and reflects the growing number of polarized nuclei that are needed to compensate the field with increasing $B_{ext}$. The sharp drop of $P_{th}$ at $B_{ext} > 1.75$ T is particularly significant. We note that when all nuclear spins in GaAs are polarized, the Overhauser field is $B_O^{max} = 5.4$T [27, 28, 37]. Therefore, the peak in $P_{th}$ at $B_{ext} \approx -B_O^{max}/3$ corresponds to the case when all spins projections, which were initially aligned randomly along the $\pm z$, are oriented along $+z$. This is the maximal nuclear polarization that can be induced by electrons. Beyond this point, the Overhauser field cannot grow further, and the dark to bright conversion is suppressed.

Let us now examine the effect of RF radiation when the system is subjected to an external magnetic field. Diagonalizing $H_{hf}$ in this case, the excitation energy is given by $\sqrt{Ng^2 + (\omega_e - \omega_n)^2}$, where $\omega_{e,n}$ are the electron and nuclear Zeeman splitting, respectively (See SI section 3). We first consider the $B_{ext} < 1.75$ T case, where $B_O = -B_{ext}$, and we can therefore set $\omega_e = 0$. In this range the resonance frequency should be given by $f_i(B) = \sqrt{f_i^2 + (\gamma_i B_{ext})^2}$. In Fig. 4c we compare the measured dependence of the $f_1$ resonance on $B_{ext}$ with the predicted behavior taking the value for $\gamma_{As}^{75}$. It is seen that there is a good qualitative and quantitative agreement.

When $B_{ext} > 1.75$T, the external field cannot be cancelled by $B_O$, and there is a net magnetic field acting on the electron. Since $\omega_e$ is much larger than both $\omega_n$ and $\sqrt{N}g$, the excitation energy of the condensate can be approximated by $\approx \hbar\omega_e$, which is in the GHz frequency range. Indeed, in this case we do not detect any measurable effect of the RF radiation on $P_{th}$ in the range $0 - 100$ MHz. This is demonstrated in Fig. 4d, which shows the relative depth of the first resonance as a function of $B_{ext}$. The abrupt disappearance of the resonance at 1.75 T is clearly visible.

**Discussion**

The observation of high frequency RF resonances, which are two orders of magnitude larger than the single exciton hyperfine resonances, and their non-trivial dependence on magnetic field, are well explained by a collective model, of $N$ electrons in the exciton BEC ground state, each coupled to a nucleus. We note that the $\sqrt{N}$ enhancement of the hyperfine coupling is similar to that found in the Tavis-Cummings model, describing the energy spectrum of $N$ spins interacting with a radiation field [40, 41]. Indeed, the two important properties of that model are present also here: The interaction Hamiltonian commutes with the total electron spin operator, and the system of $N$ spins is in its ground state. In that sense, the occurrence of these resonances confirms the collective nature of the excitons and provides important insights.

We wish to note that the fact that the nuclei are polarized throughout the mesa, and the persistence of this polarization for a long time after the optical excitation is switched off may offer an interesting system for quantum information applications using CQW or bilayer structures.

**Acknowledgments:** We wish to thank Igor Rozhansky and Lucio Frydman for fruitful discussions. This work is supported by the Israeli Science Foundation, Grant 2139/20.

**Author contributions:** A.J. and I.B.J. designed research; A.J. performed experiment; S.M. performed initial experiment; V.U. grew the high-quality molecular beam epitaxy wafer; A.J. analyzed data; M.S. developed theory; I.B.J. wrote the paper; A.J., M.S., and I.B.J. contributed to the interpretation of the results.

**Competing interests:** The authors declare no competing interest.

**Figures**

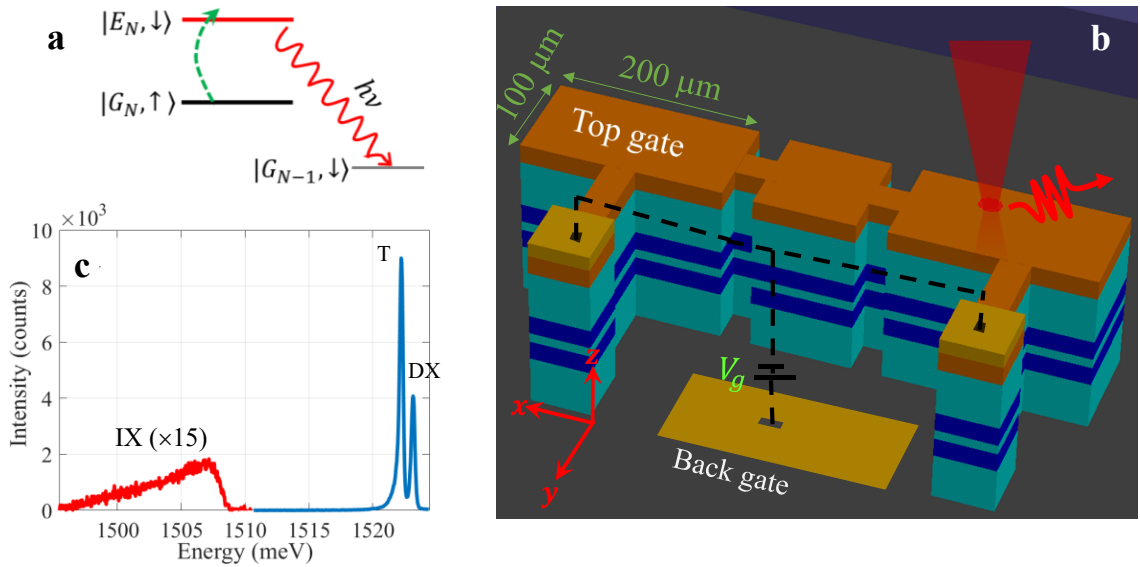

*Fig. 1. **Device and spectra**. **a**. Schematic drawing of the radiatively assisted flip-flop transition. Here N is the number of excitons. **b**. Schematic drawing of the sample structure, where the blue strips in the center represent the CQW. **c**. A typical PL spectrum at low power: DX is the direct exciton, T – the trion, and IX – the indirect exciton. Here $V_g = -4$ V, P = 0.5 μW, and T = 0.6 K.*

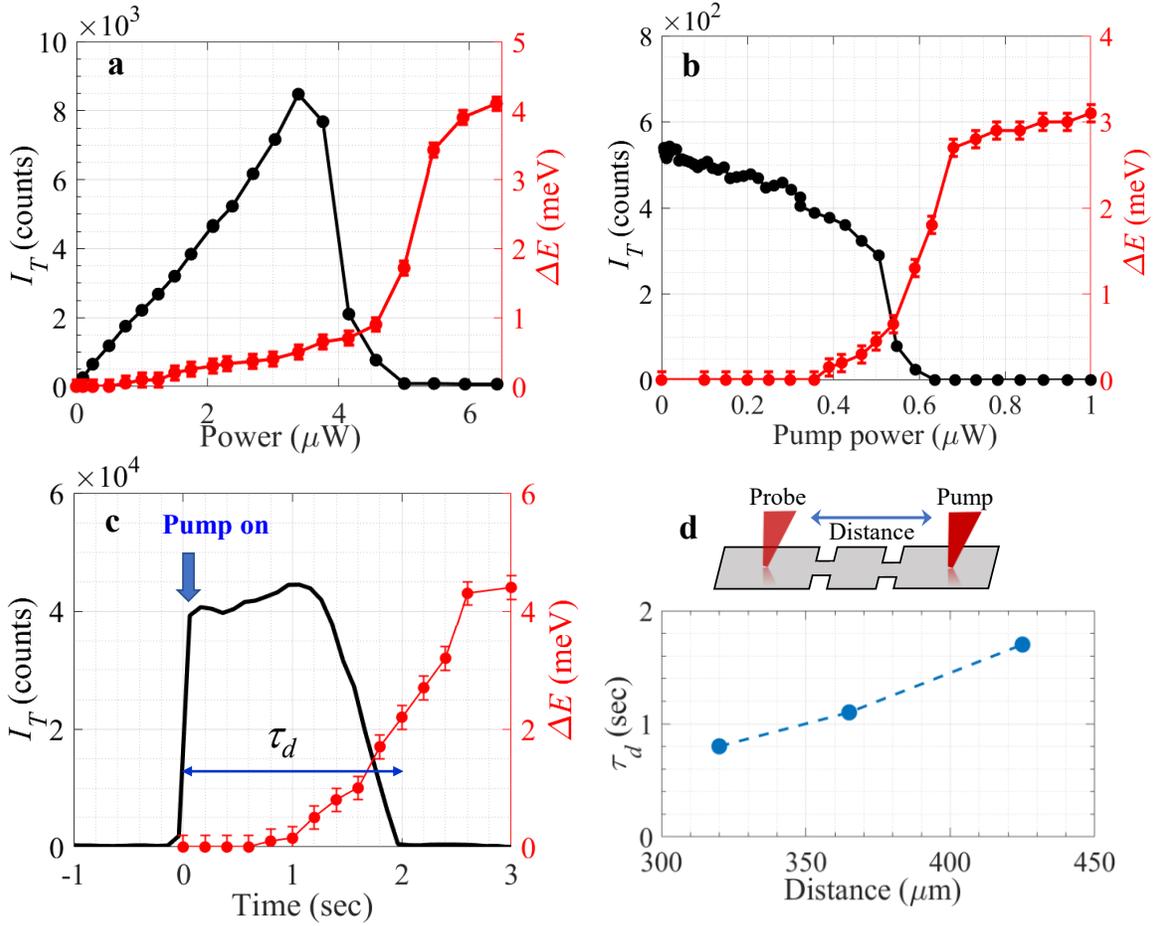

*Fig. 2. Dark exciton condensation and buildup of nuclear polarization. a. The trion line intensity, $I_T$, and the blueshift, $\Delta E$, as a function of power at 1.5 K. The threshold is clearly visible. b. $I_T$ and $\Delta E$ at the probe location as function of pump power at $T = 0.6$ K (see schematic drawing). The threshold power is lower than in (a) due to the lower temperature. c. The evolution of $I_T$ and $\Delta E$ with time at 1.5 K. The delay time, $\tau_d$, is calculated from the onset of pump, $t = 0$, to when $I_T = 0$. d. The delay time at different probe positions, where the excitation beam is fixed (origin) at 0.6 K.*

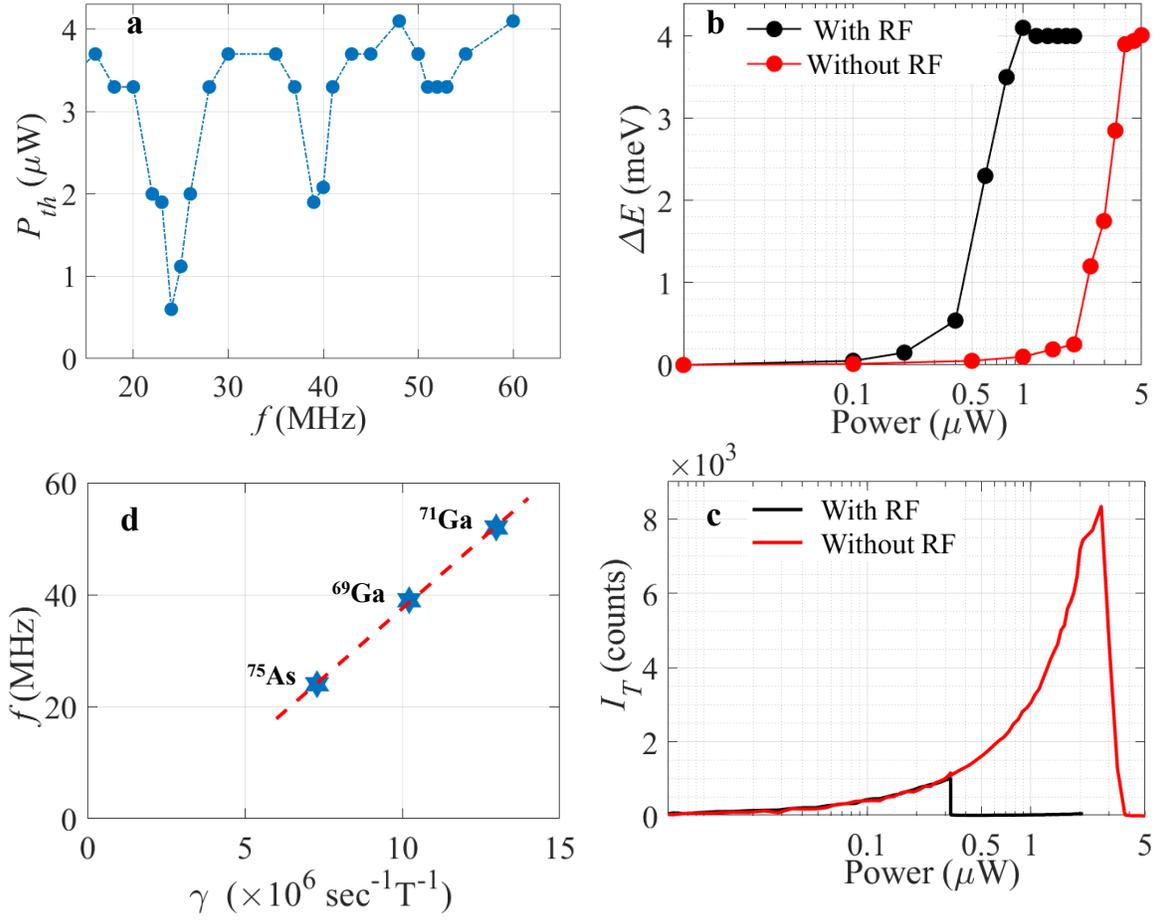

*Fig. 3. The RF resonances. a.* *The threshold power, $P_{th}$, as a function of radio frequency (RF). The three resonances at $f_i$ =24, 39 and 52 MHz are clearly seen.* *b, c*. *The blueshift, $\Delta E$, and trion intensity, $I_T$, as a function laser power without (red) and with (black) RF radiation at 24 MHz.* *d.* *The dependence of resonance frequencies, $f_i$, on gyromagnetic ratio $\gamma_i$ of the three isotopes in GaAs, $^{75}$As, $^{69}$Ga, and $^{71}$Ga. It is seen that $f_i$ are exactly proportional to $\gamma_i$.*

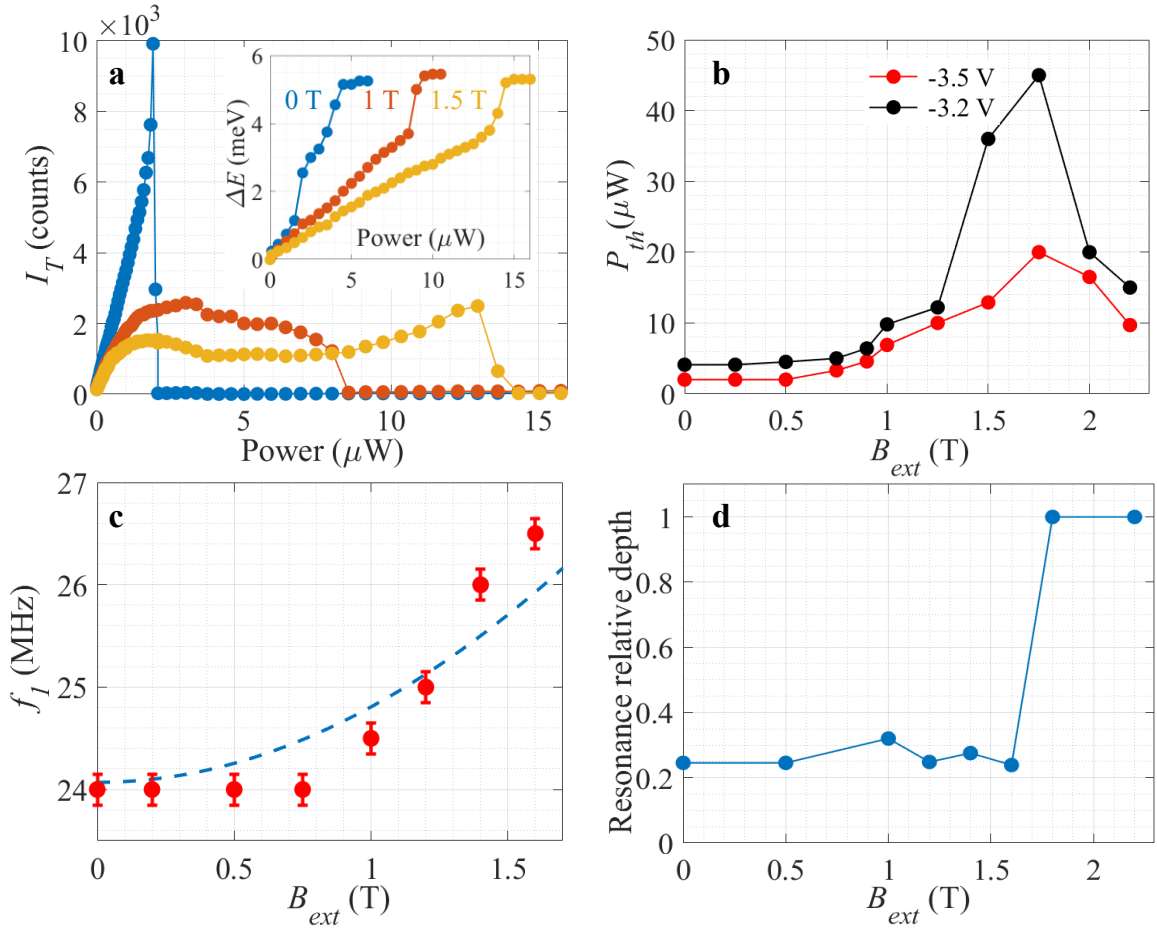

*Fig. 4. The behavior in a magnetic field $B_{ext}$. a. Trion intensity, $I_T$, and blueshift, $\Delta E$, as function of laser power at $B_{ext} = 0$ (blue), 1 T (orange), and 1.5 T (yellow). $V_g = -3.5$ V and T = 1.5 K. The intermediate range, at which $I_T$ is approximately constant, is clearly seen. b. $P_{th}$ as a function of magnetic field for two different gate voltage at T= 1.5 K. c. The $^{75}$As resonance, $f_1$, as a function of magnetic field. The dashed line shows the predicted behavior. d. The resonance depth, $P_{th}(f_1)/P_{th}(0)$, as a function of magnetic field at T= 1.5 K. The resonance disappears at $B_{ext} > 1.75$ T.*

# Supplementary Information & Extended Data

# Giant Hyperfine Interaction between a Dark Exciton Condensate and Nuclei

A. Jash[1], M. Stern[2], S. Misra[1], V. Umansky[1], and I. Bar Joseph[1]

[1]Department of Condensed Matter physics, Weizmann Institute of Science, Rehovot 7610001, Israel
[2] Department of Physics, Bar-Ilan University, Ramat-Gan 5290002, Israel

# Contents

# 1. Experimental Details and Methods

## a. Sample Structures

The coupled quantum wells (CQW) consist of two quantum wells having width 12 and 18 nm and a barrier of 3 nm $Al_{0.28}Ga_{0.72}As$. The CQW is embedded between two superlattices (SL) of thickness ~1 μm each. The SL below the CQW has 33 periods of [27 nm $Al_{0.37}Ga_{0.63}As$ + 2 nm AlAs + 1 nm GaAs], and above the CQW has 20 periods of [50 nm $Al_{0.37}Ga_{0.63}As$ + 1 nm GaAs]. The top (0.15 μm) and bottom layers (0.25 μm) are silicon doped ($n = 10^{18}$ cm$^{-3}$) $Al_{0.12}Ga_{0.88}As$. The band gap of the contact layers is 1.650 eV, whereas for the SL layers, it is ~ 1.9 eV. These values are well above the excitation laser energy, so that these layers do not contribute to carrier photo-generation.

Mesas of various shapes and sizes were prepared using optical lithography and wet etching after the MBE growth of the sample. Most of the data presented in this work was collected using an elongated structure, which is depicted in Fig. 1 of the manuscript.

## b. Photoluminescence Measurements Setup

The experiments were conducted using two different cryostats. A dilution refrigerator with optical access, which allows conducting PL measurements in the range of 100 mK to 5 K, and a split-coil magneto optical pumped-helium cryostat, with a temperature range of 1.5 K to 10 K, and magnetic fields up to 7T. Throughout the measurements, we explored a temperature range of 0.1 K to 6 K and a magnetic field range of 0 to ± 5 T. Thermal cycling between 4 K and 300 K was performed multiple times during the measurements. Notably, no changes in the optical properties were observed.

For the single beam measurements, we illuminated the sample using a Ti: sapphire laser with a Gaussian spot size of $\sigma = 20$ μm and an energy of 1.5287 eV. We have adjusted the laser energy to be below the energy gap of the narrow well such that carriers are created in the wide well only. In the pump-probe experiment, a diode laser with the same Gaussian spot size and an energy of 1.5794 eV was used as the probe beam.

To collect photoluminescence (PL) from different regions of the mesa, a pinhole with a diameter of 50 μm was positioned at the objective's focal plane. The PL from the sample was guided to a Spex (500M) spectrometer and imaged and analyzed using an Andor iXon camera. A Keithley 2400 source meter was used to apply the gate voltage. LabVIEW and Andor software are used for the acquisition and recording of all the experimental measurements.

### c. RF Measurements Setup

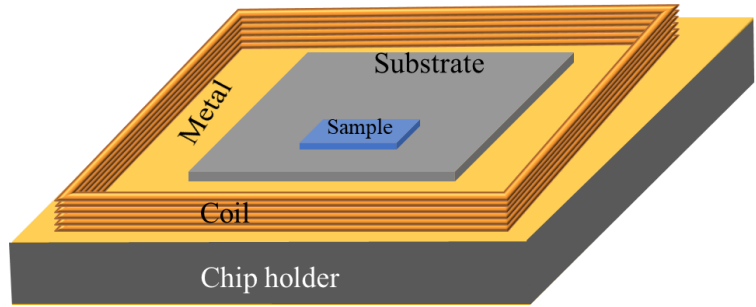

*Figure S1. A schematic drawing of the RF measurements setup.*

The configuration, which was implemented to generate an oscillating magnetic for the RF measurements, is described in Fig. S1. The coil, which has six layers with five turns each, was wound on a bobbin prepared using a 3D printer, employing PVA material. Once the coil was wound, the PVA material was dissolved by water, and the coil was fixed to a 24-pin chip header with epoxy. We note that the metal surface of the chip holder strongly suppresses the z component of the RF field at the area near the center of the coil. Hence, by placing the mesa a little off-center, we obtain a component of the RF field in the CQW plane. Indeed, we verified that similar results were obtained when the coil was oriented in a perpendicular direction.

To generate RF magnetic field, we employed a function generator from Berkeley Nucleonics Corp (Model A2255), which has a frequency range of 1-250 MHz and an adjustable amplitude of 1-10 V. Throughout the RF measurement process, we took caution to prevent excessive heating of the sample. Indeed, we did not observe any such effects. The current flowing through the coil remained in the range of a few milliamperes, resulting in RF radiation at the sample surface within the nanowatt range.

To determine $P_{th}$ at a particular frequency, we fixed the RF radiation at a frequency within the $0 - 100$ MHz range and measured the PL spectrum as the power is ramped up at a slow rate, waiting 5 seconds between consecutive measurements. To obtain a curve like Fig. 3a we repeated this process in steps of 0.5 MHz over a broad RF range.

## 2. Diffusion Model

To model the dark condensate – nuclear interaction we conducted a simple one-dimensional diffusion simulation of the following couples equations for the dark exciton density, $n$, and polarized nuclear density, $N$

$$\frac{\partial n}{\partial t} = P - \frac{n}{\tau_{nr}} + D\frac{\partial^2 n}{\partial x^2} - \alpha n(N_0 - N)$$

$$\frac{\partial N}{\partial t} = -\frac{N}{\tau_N} + \alpha n(N_0 - N)$$

Here $P = P_0 \exp\left(-\frac{x^2}{\sigma^2}\right)$ is the excitation pump beam, which is assumed to have a gaussian profile, $\tau_{nr}$ is the non-radiative recombination time of dark excitons, $D$ is the exciton diffusion constant, $N_0$ is the total nuclear density, $\tau_N$ is the nuclear relaxation time, and $\alpha$ is the coupling term, which polarizes the nuclei and causes dark exciton to disappear by turning it into bright.

In Fig. S2 we show the time evolution of the exciton density, $n$, and nuclear polarization, $N$, at the pump location. The delay time, $\tau_d$, which characterized the buildup of in $n$ and $N$ is clearly seen.

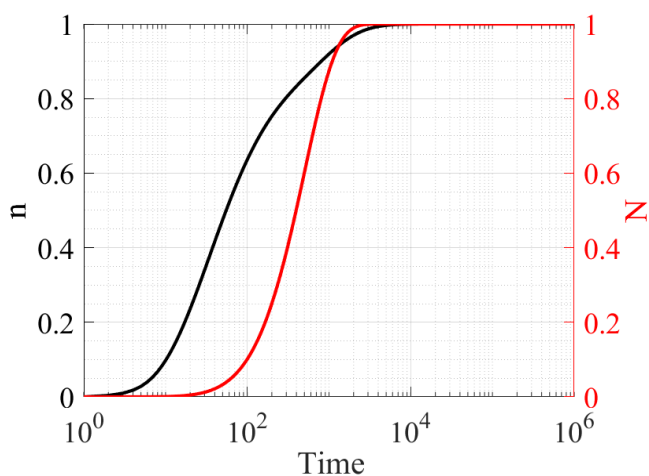

*Figure S2. The evolution of exciton density and nuclear polarization with time.*

# Hyperfine Coupling of Nuclei with a Condensate

### a. Hyperfine interaction between single nucleus and single electron

Let's first consider a single nucleus coupled to a single electronic spin by hyperfine interaction. The Hamiltonian of the system is given by,

$$H = \frac{1}{2} \hbar \omega_n \sigma_{z,n} + \frac{1}{2} \hbar \omega_e \sigma_{z,n} + \hbar g/4 \sigma_e \cdot \sigma_n$$

Here $\sigma_i$ are the Pauli matrices, $g$ is the hyperfine constant, and $\omega_{e,n}$ are the cyclotron frequency of electron and nucleus, respectively, in a magnetic field $B$.

This Hamiltonian can be written as

$$H = \frac{1}{2} \hbar \omega_n (I \otimes \sigma_{z,n}) + \frac{1}{2} \hbar \omega_e (\sigma_{z,e} \otimes I) + \frac{1}{4} \hbar g (\sigma_{z,e} \otimes \sigma_{z,n}) + \frac{1}{2} \hbar g (\sigma_e^+ \sigma_n^- + \sigma_e^- \sigma_n^+)$$

$I$ is unit matrix Denoting the electron and nuclear states by $\pm$ and $\uparrow, \downarrow$, respectively, the Hamiltonian can be represented in the basis $|+,\uparrow>, |+,\downarrow>, |-,\uparrow>,$ and $|-,\downarrow>$ as

$$H = \frac{\hbar}{2} \begin{pmatrix} \omega_e + \omega_n + g/2 & 0 & 0 & 0 \\ 0 & \omega_e - \omega_n - g/2 & g & 0 \\ 0 & g & -\omega_e + \omega_n - g/2 & 0 \\ 0 & 0 & 0 & -\omega_e - \omega_n + g/2 \end{pmatrix}$$

The eigenvalues of this Hamiltonian are

$$E_{+\uparrow} = \frac{\hbar}{2}\left[\omega_e + \omega_n + \frac{g}{2}\right]$$

$$E_{+\downarrow} = \frac{\hbar}{2}\left[\sqrt{(\omega_e - \omega_n)^2 + g^2} - \frac{g}{2}\right]$$

$$E_{-\uparrow} = \frac{\hbar}{2}\left[-\sqrt{(\omega_e - \omega_n)^2 + g^2} - \frac{g}{2}\right]$$

$$E_{-\downarrow} = \frac{\hbar}{2}\left[-\omega_e - \omega_n + \frac{g}{2}\right]$$

The NMR transitions correspond to

$$\Delta E_1 = E_{+\uparrow} - E_{+\downarrow} \simeq \hbar(\omega_n + g/2)$$

$$\Delta E_2 = E_{-\uparrow} - E_{-\downarrow} \simeq \hbar(\omega_n - g/2)$$

The ESR transitions correspond to

$$\Delta E_3 = E_{+\uparrow} - E_{-\uparrow} \simeq \hbar(\omega_e + g/2)$$

$$\Delta E_3 = E_{+\downarrow} - E_{-\downarrow} \simeq \hbar(\omega_e - g/2)$$

where we took into account that $\omega_e - \omega_n \gg g$.

### b. Hyperfine interaction between a nuclear spin and with the electrons of an exciton BEC

Using Eq. (1), one can write the Hamiltonian of the system

$$H = \frac{1}{2}\hbar\omega_n(I \otimes \sigma_{z,n}) + \frac{1}{2}\hbar\omega_e(\Sigma\sigma_{z,e}^{(i)} \otimes I) + \frac{1}{4}\hbar g(\Sigma\sigma_{z,e}^{(i)} \otimes \sigma_{z,n}) + \frac{1}{2}\hbar g\Sigma(\sigma_{e,(i)}^{+}\sigma_n^{-} + \sigma_{e,(i)}^{-}\sigma_n^{+})$$

Using the total electron spin operator $S = \Sigma\sigma_e^{(i)}$ the Hamiltonian can be written as

$$H = \frac{1}{2}\hbar\omega_n\sigma_{z,n} + \frac{1}{2}\hbar\omega_e S_z + \frac{1}{4}\hbar g S_z \otimes \sigma_{z,n} + \frac{1}{2}\hbar g(S_+\sigma_{-n} + S_-\sigma_{+n})$$

The electronic system is supposed to be initially in its ground state $|G\rangle = |-,-,-,\ldots-\rangle$ of N electrons. The first excited state of the system consists of a single electron spin flip and can be written as $|E\rangle = 1/\sqrt{N}(|+,-,-,\ldots-\rangle + |-,+,-,\ldots-\rangle + \cdots |-,-,-,\ldots+\rangle)$ [1].

We can represent the Hamiltonian of the system in the subspace $|G,\uparrow\rangle, |G,\downarrow\rangle, |E,\uparrow\rangle$, and $|E,\downarrow\rangle$ as 4×4 matrix with out of the diagonal matrix element

$$\langle E,\downarrow|H|G,\uparrow\rangle = \frac{1}{\sqrt{N}}\frac{\Sigma\hbar g}{2} = \sqrt{N}\,\hbar g/2$$

And thus

$$H = \frac{\hbar}{2}\begin{pmatrix} \omega_e + \omega_n + g/2 & 0 & 0 & 0 \\ 0 & \omega_e - \omega_n - g/2 & g\sqrt{N} & 0 \\ 0 & g\sqrt{N} & -\omega_e + \omega_n - g/2 & 0 \\ 0 & 0 & 0 & -\omega_e - \omega_n + g/2 \end{pmatrix}$$

with eigen energies:

$$E_{E\uparrow} = \frac{\hbar}{2}\left[\omega_e + \omega_n + \frac{g}{2}\right]$$

$$E_{E\downarrow} = \frac{\hbar}{2}\left[\sqrt{(\omega_e - \omega_n)^2 + Ng^2} - \frac{g}{2}\right]$$

$$E_{G\uparrow} = \frac{\hbar}{2}\left[-\sqrt{(\omega_e - \omega_n)^2 + Ng^2} - \frac{g}{2}\right]$$

$$E_{G\downarrow} = \frac{\hbar}{2}\left[-\omega_e - \omega_n + \frac{g}{2}\right]$$

Note that we kept the labelling of the energy levels, even though the hyperfine interaction mixes the states $|G,\uparrow\rangle$ and $|E,\downarrow\rangle$. The energy difference between $E_{E\downarrow}$ and $E_{G\uparrow}$ is:

$$\Delta E = E_{E\downarrow} - E_{G\uparrow} = \hbar\left[\sqrt{(\omega_e - \omega_n)^2 + Ng^2}\right]$$

At zero magnetic field it becomes $\Delta E = \hbar\sqrt{N}g$.

We wish to comment here that the linear fit of Fig. 3d does not extrapolate to zero at $\gamma = 0$. This implies that the measured resonances are shifted by $-12$ MHz, independent of the isotope involved. We do not know the reason for this shift.

# 3. Extended Data

### a. Gate voltage dependence

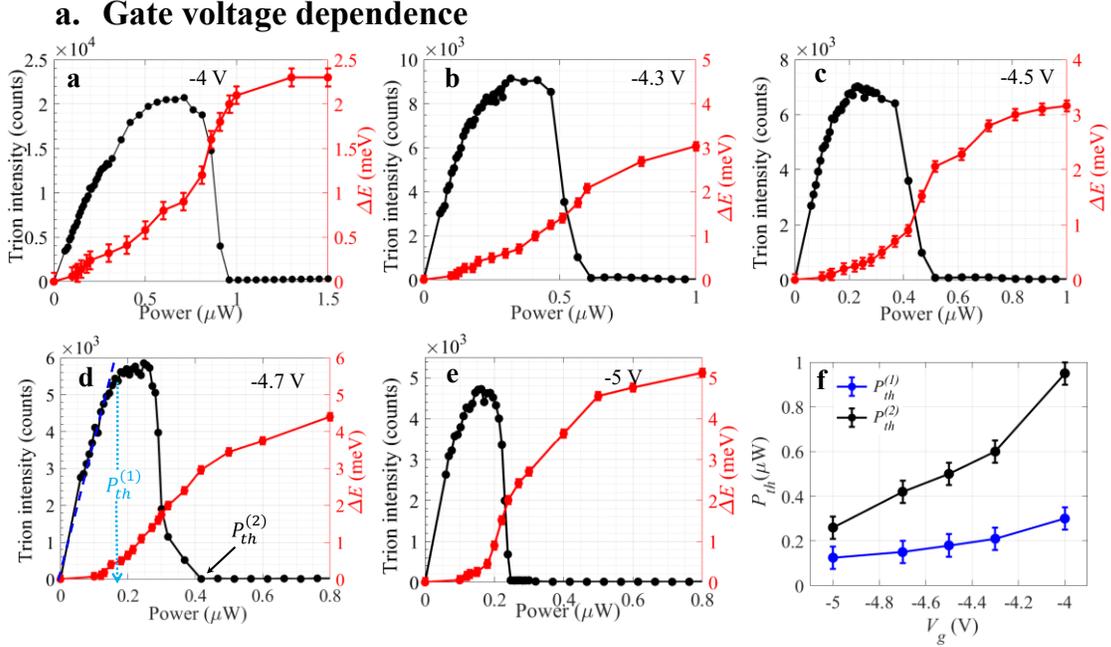

*Figure S3. (a-e) The trion intensity (left axis) and blueshift (right axis) as function power at 0.6 K at different gate voltage. In Fig. (d), the blue line depicts the linear relationship between trion intensity and power. The extraction of threshold power, $P_{th}^{(1)}$ and $P_{th}^{(2)}$, from the data, using two methods is presented in Fig. (d). (f) The measured $P_{th}$ for the two methods as a function of gate voltage at 0.6 K.*

Figure S3(a-e) illustrates the trion intensity, $I_T$, and blueshift, $\Delta E$, as functions of laser power for various gate voltages at 0.6 K. The electrons from the wide well (WW) tunnel to the narrow well (NW) under the influence of gate voltage. The tunneling time closely compares with the direct exciton (DX) formation time, causing some photoexcited electrons to remain in the WW and form direct excitons. Excess holes in the WW have two options: they can either bind to the electrons in the NW, forming indirect excitons (IX), or bind to the excitons, forming trions (T). At very low power, below 100 nW, the favorable condition for holes is to bind to excitons and form trions rather than forming IX with NW electrons. At a certain threshold power, labeled as $P_{th}^{(1)}$ in Fig. S3(d), the trion intensity, $I_T$, deviates from linear increase with power and saturates. At this point, holes start binding with the NW electrons and the IX density increases. This is manifested as a blueshift of the IX energy, $\Delta E$. Simultaneously, the trion intensity drops and eventually reaches zero. We label the power at which $I_T$ becomes zero as $P_{th}^{(2)}$. In Fig. S3(f) we show the gate voltage dependence of the two threshold powers. It is evident that the two curves differ by a constant scaling factor. Since the point at which $I_T = 0$ can be easily and unambiguously determined, we chose it as the definition of $P_{th}$.

The two methods of extracting threshold yield slightly different critical condensation density, $n_c$. Using $P_{th}^{(1)}$ we obtain $\Delta E = 0.4$ meV at 0.6K, which for uncorrelated excitons corresponds to $n_c \approx 1.6 \times 10^9$ cm$^{-2}$. This value is in good agreement with the theoretical value, $n_c = \frac{M}{h^2} k_B T$, which gives $n_c \approx 2 \times 10^9$ cm$^{-2}$.

### b. Temperature dependence

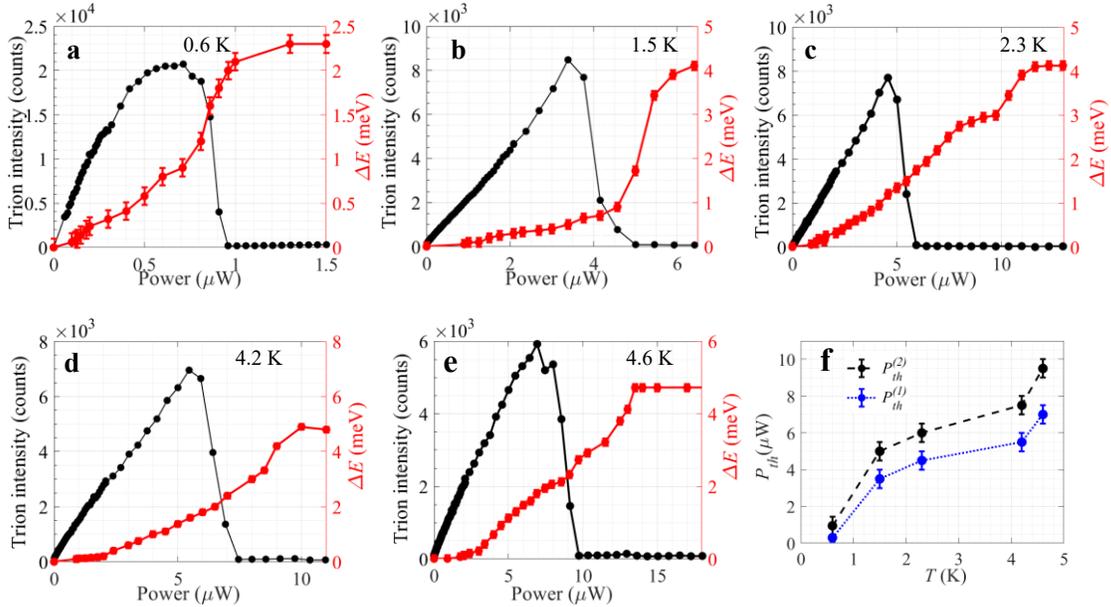

*Figure S4. (a-e) The trion intensity and blueshift as function of laser power at $V_g = -4$ V for different temperature. (f) $P_{th}^{(1)}$ and $P_{th}^{(2)}$ as a function of temperature. (f) The measured $P_{th}$ for the two methods at - 4 V.*

Figure S4(a-e) displays the trion intensity and blueshift as a function of excitation power at four different temperatures: 0.6 K, 1.5 K, 2.3 K, 4.2 K, and 4.6 K. Note that the measurements at 0.6 K were performed with improved collection system, yielding a 10-

fold increase in the PL intensity). It is clearly seen that $P_{th}$ increases with temperature, and this is manifested both in $P_{th}^{(1)}$ and in $P_{th}^{(2)}$.

### c. Condensation density

The critical density for condensation, $n_c$, is expected to increase linearly with temperature, $n_c = \frac{M}{h^2} k_B T$, where $k_B$ and $h$ are the Boltzmann and Planck constants, respectively, and $M$ is the exciton mass. Indeed, we find that $\Delta E$ at threshold, which is proportional to $n_c$, increases linearly with temperature (Fig. S5(a)) and is nearly independent of gate voltage at constant temperature (Fig. S5(b)).

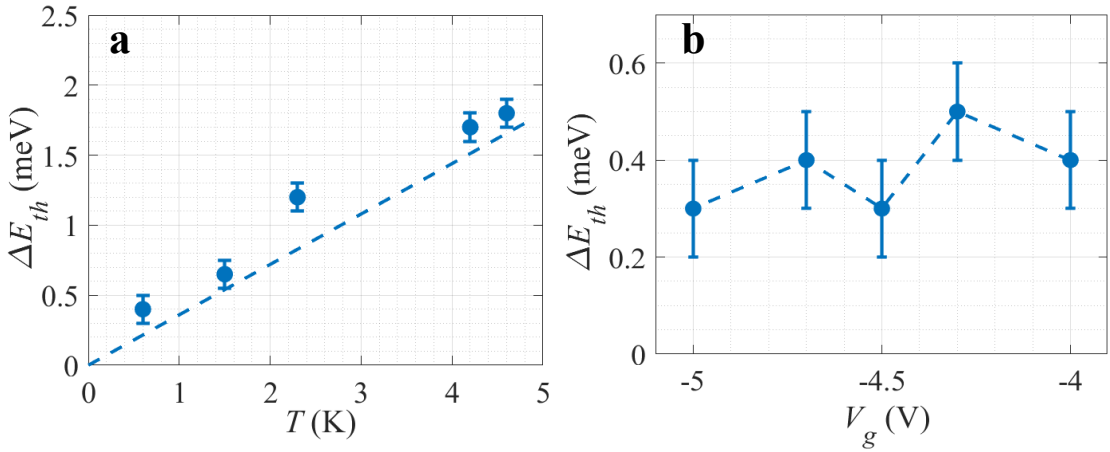

*Fig. S5. The threshold blueshift, $\Delta E_{th}$, as function of (a) temperature at $V_g = -4.0$ V and (b) gate voltage at 0.6 K. The points are taken where trion intensity deviates from linear behaviour ($P_{th}^{(1)}$).*

### d. Slow build-up of the dark excitons density

Figure S6(a-c) presents the time evolution of the spectrum for various laser powers at $T = 1.5$ K and $V_g = -4$ V, following switching on the excitation laser at $t = 0$. The two high energy lines are the direct exciton (DX) and trion (T). The weak IX line is outlined by white dashed line. It is seen that after a delay time ($\tau_d$) of a few seconds, which decreases with increasing power, the trion line vanishes, and the IX energy undergoes an abrupt blueshift. It should be noted that at later times, the intensity of the DX line diminishes, and a new line labelled as Z emerges at a lower energy and eventually becomes the dominant line [3,4]. In this study, we focus on the power range below the emergence of the Z line.

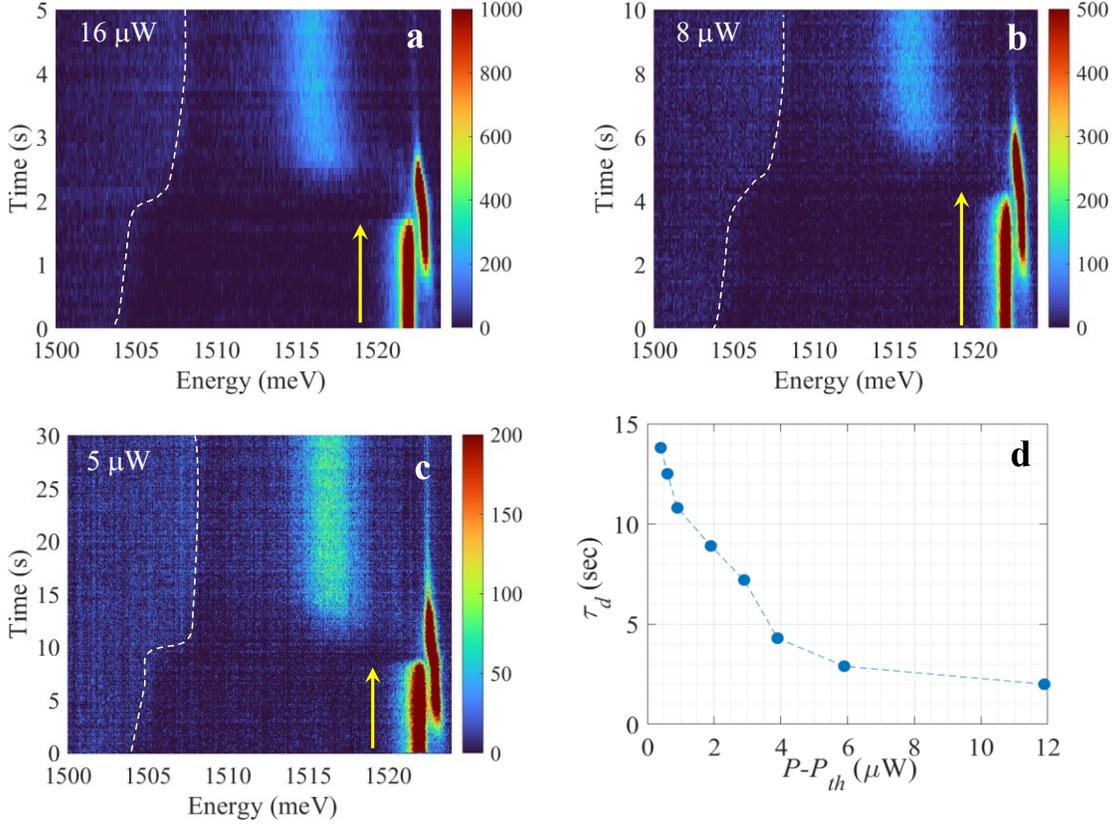

Figure S6. (a-c) The time evolution of the spectrum for various laser powers at T = 1.5 K and $V_g = -4$ V. The IX energy is outlined by white dashed line. (d) The delay time ($\tau_d$) as function of excess pump power.

### e. The decay of nuclear polarization

To determine the decay of the nuclear polarization, we conduct pump-probe measurements, where we follow the recovery of the spectrum at the probe position following an abrupt switching off the pump power. The experiment was conducted using a 5 µW pump and 50 nW probe ($P_{th} = 0.9$ µW). We first turned on the pump for 50 seconds, letting the sample to reach its steady state, and then turned it off at $t = 0$ (as depicted in Fig. S7). We find that the trion line intensity ($I_T$) began to increase approximately 3 seconds after the pump was turned off and reached its saturated value after more than ten seconds. The slow recovery of the trion line marks the decay of nuclear polarization at the probe position.

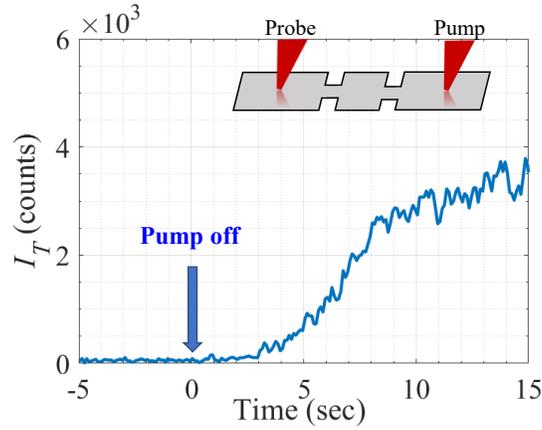

*Figure S7. The time evolution of the trion line intensity, $I_T$, at the probe location after the pump is switched off at $t = 0$. Here $T = 0.6$ K and $V_g = -4$ V.*

**f.  RF measurements**

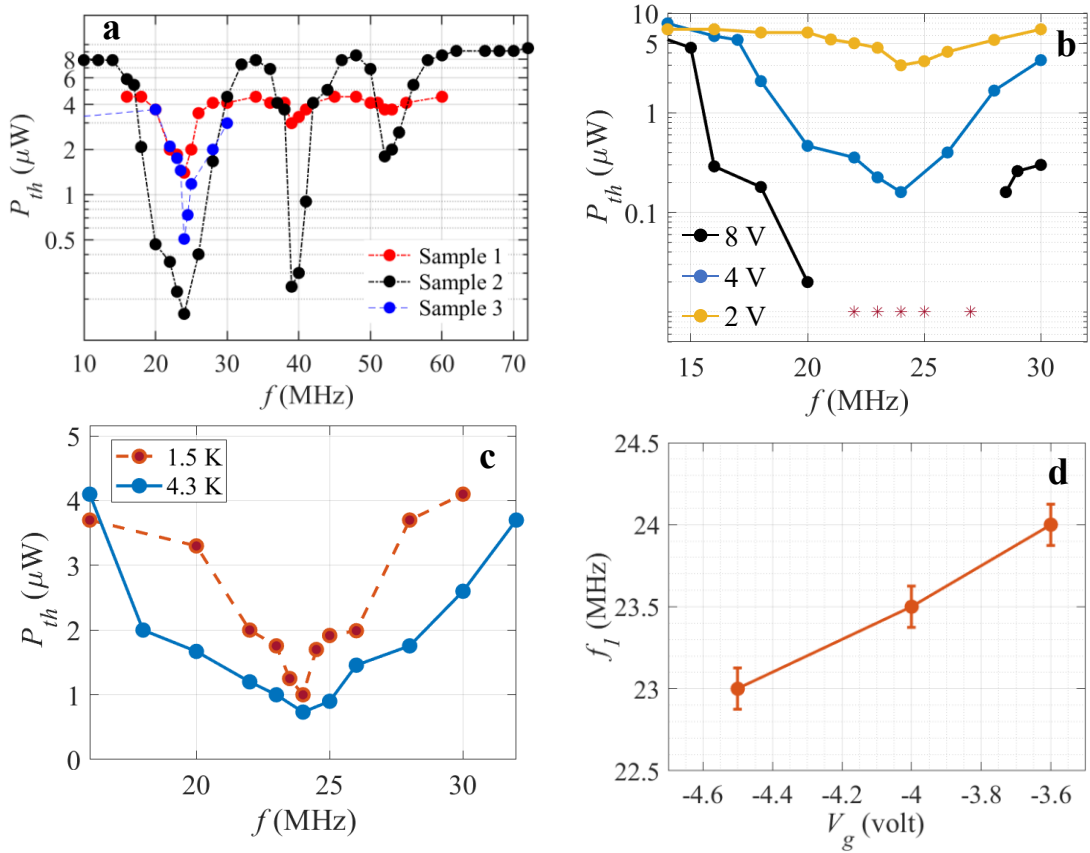

*Figure S8. The measured $P_{th}$ as function of RF frequency at T=1.5K and $V_g$ = -4V; (a) at three different samples; (b) at three RF voltages in sample 2; (c) at two different temperatures with a fixed $V_g$ of - 4 V. (d) The first resonance different with gate voltage at 1.5 K.*

Figure S8 illustrates the variation the RF resonances as we change various parameters. In (a) we show the RF resonances for three different mesas. It is evident that while the resonance depth varies, the resonance frequency is the same. In (b) we show how $P_{th}$ becomes lower at resonance with increasing RF power. At the highest RF voltage in the range $20 - 27$ MHz, $P_{th}$ is lower than 10 nW and could not be determined accurately (shown by star symbol). In (c) we show the resonance in two temperatures, 1.5 and 4.3 K. We can see that the resonance is broadened at higher temperature, but the resonance frequency remains the same. Finally, in (d) we demonstrate the dependence of the resonance frequency on gate voltage. We notice a small shift of the resonance to lower frequency as $|V_g|$ increases. We believe that this reflects the change of the electron wavefunction, $\psi$, which is pushed into the barrier and thereby reducing the factor $|\psi(0)|^2$ in the hyperfine coupling.

### g. Voigt and Faraday configurations

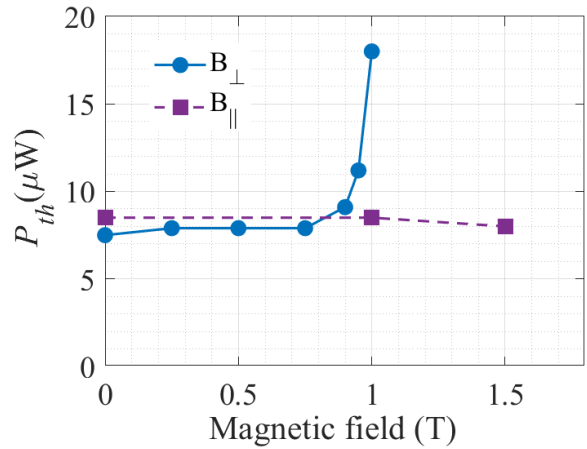

*Figure S9. $P_{th}$ as function of magnetic field for Voigt, $B_\parallel$, and Faraday, $B_\perp$, configurations. Here T = 1.5 K and $V_g = -4$ V (measurements were performed on sample 2).*

Figure S9 compares the measured $P_{th}$ values in Voigt and Faraday configurations (magnetic field aligned parallel or perpendicular to the quantum well plane, respectively).

It is evident that while there is a strong dependence of $P_{th}$ on magnetic field in Faraday configuration, no substantial variations in $P_{th}$ is observed in response to magnetic fields up to $B_\parallel = 1.5$ T.